\documentstyle[mathrsfs,pra,aps]{revtex}
\begin{document}
\draft

\def\overlay#1#2{\setbox0=\hbox{#1}\setbox1=\hbox to \wd0{\hss #2\hss}#1%
\hskip
-2\wd0\copy1}
\twocolumn[
\hsize\textwidth\columnwidth\hsize\csname@twocolumnfalse\endcsname

\title{Space-time geometry of quantum dielectrics}
\author{Ulf Leonhardt}
\address{School of Physics and Astronomy, University of St Andrews, 
North Haugh, St Andrews, Fife, KY16 9SS, Scotland}
\address{Physics Department, Royal Institute of Technology (KTH),
Lindstedtsv\"agen 24, S-10044 Stockholm, Sweden}
\maketitle
\begin{abstract}
Light experiences dielectric matter as an effective gravitational field
and matter experiences light as a form of gravity as well. 
Light and matter waves see each other as dual space--time metrics, 
thus establishing a unique model in field theory.
{\it Actio et reactio} are governed by Abraham's energy--momentum
tensor and equations of state for quantum dielectrics.
\end{abstract}
\date{today}
\vskip2pc
]
\narrowtext

\section{Introduction}

A moving dielectric medium appears to light as an effective
gravitational field \cite{Gordon,PhamMauQuan,LPstor,LPliten}.
The medium alters the way in which an electromagnetic field perceives
space and time, formulated most concisely in Gordon's effective
space--time metric \cite{Gordon}
\begin{equation}
\label{gordon}
g^F_{\alpha\beta} = g_{\alpha\beta} + 
\left(\frac{1}{\varepsilon\mu} - 1 \right)u_\alpha u_\beta
\,\,.
\end{equation}
We allow for a back--ground metric $g_{\alpha\beta}$,
mostly to have the convenience of choosing arbitrary coordinates,
but also for the possible inclusion of a genuine gravitational field. 
Gordon's metric (\ref{gordon}) depends on the dielectric properties of
the medium, on the permittivity $\varepsilon$ and on the
magnetic permeability $\mu$, as well as on the
four--dimensional flow $u^\alpha$ of the medium 
(the local four--velocity).
The product $\varepsilon\mu$ is the square of the refractive index and
the prefactor $1-(\varepsilon\mu)^{-1}$ is known as Fresnel's dragging
coefficient \cite{Moeller,Fresnel,Fizeau} (in Fresnel's days
the part of the ether that the moving medium is able to drag 
\cite{Fresnel}). 
In the limit of geometrical optics \cite{BornWolf}, light rays are
zero--geodesic lines with respect to Gordon's metric 
\cite{Gordon,PhamMauQuan,LPstor,LPliten}.
In the special case of a medium at rest, this result is equivalent to
Fermat's principle \cite{BornWolf} and to the formulation of
geometrical optics as a non--Euclidean geometry in space
\cite{Bortolotti}.

Light sees dielectric matter as an effective space--time metric. 
How does matter see light? 
In atom optics \cite{atom_optics}, the traditional role of light and
matter is reversed: Atomic de--Broglie waves are subject to atom--optical 
instruments made of light. 
Light acts on matter waves in a similar way as matter acts on light. 
This paper indicates that an atomic matter wave experiences an
electromagnetic field as the effective metric 
\begin{equation}
\label{metric}
g^A_{\alpha\beta} = \left( 1 - a\, {\mathscr L}_F \right) 
g_{\alpha\beta} - b\,T^F_{\alpha\beta}
\end{equation}
with
\begin{equation}
\label{ab}
a = \frac{1}{mc^2\rho} \left(\varepsilon + \frac{1}{\mu} - 2 \right)
\quad,\quad
b = \frac{1}{mc^2\rho} \left(\varepsilon - \frac{1}{\mu} \right)
\,\,.
\end{equation}
Here ${\mathscr L}_F$ is the Lagrangian of the free electromagnetic
field, defined in Eq.\ (\ref{ldef}), and $T^F_{\alpha\beta}$ is the 
free--field energy--momentum tensor (\ref{tdef}).
As usual, $c$ denotes the speed of light in vacuum and
$m$ is the mass of a single dielectric atom.
In the definition (\ref{ab}), $\rho$ can be regarded as the probability
density of the atomic de--Broglie wave, for most practical purposes.
(Strictly speaking, $mc^2\rho$ describes the total enthalpy density of
the matter wave, including the rest energy as the lion's share.)
Throughout this paper we employ SI units and use the 
Landau--Lifshitz convention \cite{LL2} of general relativity (with the
exception of using greek space--time and latin space indices).
To derive the result (\ref{metric}) with the dielectric parameters
(\ref{ab}) we postulate that the interaction between light and matter
takes on the general form of a metric.
Then we demonstrate the consistency of this idea with previous
knowledge, and in particular with Gordon's metric (\ref{gordon}).

The metric (\ref{metric}) indicates that the energy--momentum of light
curves directly the space--time of a dielectric matter wave.
Under normal circumstances the deviation from the back--ground
geometry is very small, see Eqs.\ (\ref{metric}) and (\ref{ab}),
because the ratio between the electromagnetic energy and the atomic 
rest energy $mc^2$ is typically an extremely small number.
In the Newtonian limit of general relativity \cite{LL2}, the
gravitational correction to a flat Minkowski space--time is tiny as
well, because the correction is proportional to the ratio between the
potential energy and $m c^2$ of a test particle.
For weak gravitational fields and low test--particle velocities,
general relativity is an equivalent formulation of Newtonian physics
that agrees in all predicted effects and yet establishes a radically
different physical interpretation.
Similarly, given the current state of the art in atom optics, the idea
that light curves the space--time for matter waves is an equivalent 
formulation of the known light forces, i.e. of the dipole force 
and of the recently investigated R{\"o}ntgen interaction \cite{Rpapers}.
However, one can conceive of significantly enhancing the dielectric
properties of matter waves \cite{LPliten} using similar methods as in
the spectacular demonstrations of slow light \cite{slowlight}. 
Loosely speaking, a large effective dielectric constant $\varepsilon$ 
could counteract the rest energy $mc^2$ in the relations (\ref{ab}).
In this way one could use light to build atom--optical analogues of
astronomical objects on Earth, for example a black hole made of light.

\section{Electromagnetic fields}

\subsection{Field tensors}

Let us first agree on the definitions of the principal electromagnetic
quantities in SI units in general relativity. 
We employ the space--time coordinates $x^\mu = (ct,{\bf x})$.
The electromagnetic four--potential is
\begin{equation}
A_\nu = \left(U, -c {\bf A} \right)
\,\,.
\end{equation}
The electromagnetic field--strength tensor is constructed as
\begin{equation}
\label{fdef}
F_{\mu\nu} \equiv D_\mu A_\nu - D_\nu A_\mu = 
\partial_\mu A_\nu - \partial_\nu A_\mu
\,\,
\end{equation}
using the covariant derivatives $D_\mu$ with respect to the
back--ground metric $g_{\mu\nu}$.
As is well known \cite{LL2}, in the definition (\ref{fdef}) of 
$F_{\mu\nu}$ on a possibly curved space--time,
we have been able to replace the $D_\mu$
by ordinary partial derivatives 
$\partial_\mu \equiv \partial/\partial x^\mu$.
The field--strength tensor reads in local--galilean coordinates
(in a local Minkowski frame)
\begin{equation}
\label{fmn}
F_{\mu\nu} =
\left(
\begin{array}{cccc}
0 & E_x & E_y & E_z \\
-E_x & 0 & -cB_z & cB_y \\
-E_y & cB_z & 0 & -cB_x \\
-E_z & -cB_y & cB_x & 0
\end{array}
\right)
\,\,.
\end{equation}
It will become useful at a later stage of this enterprise to introduce
a four--dimensional formulation $H^{\mu\nu}$ of the dielectric 
${\bf D}$ and ${\bf H}$ fields,
\begin{equation}
\label{hmn}
H^{\mu\nu} =
\left(
\begin{array}{cccc}
0 & -D_x & -D_y & -D_z \\
D_x & 0 & -H_z/c & H_y/c \\
D_y & H_z/c & 0 & -H_x/c \\
D_z & -H_y/c & H_x/c & 0
\end{array}
\right)
\,\,,
\end{equation}
here defined in local--galilean coordinates.

\subsection{Quadratic field tensors}

In dielectric media, induced atomic dipoles constitute an interaction
between light and matter that is quadratic in the electromagnetic
field--strength tensor \cite{LL8}.
Let us therefore list a set of linearly independent second--rank
tensors that are quadratic in $F_{\mu\nu}$.
The most elementary one is the product of the metric tensor $g_{\mu\nu}$ 
with the scalar Lagrangian ${\mathscr L}_F$ of the free
electromagnetic field \cite{LL2}.
This Lagrangian is
\begin{equation}
\label{ldef}
{\mathscr L}_F = 
- \frac{\varepsilon_0}{4}\, F_{\alpha\beta} F^{\alpha\beta} =
- \frac{\varepsilon_0}{4}\, g^{\alpha \alpha'} g^{\beta \beta`}
F_{\alpha\beta} F_{\alpha'\beta'} 
\,\,,
\end{equation}
or, in local--galilean coordinates,
\begin{equation}
{\mathscr L}_F = 
\frac{\varepsilon_0}{2}\,\left( E^2 - c^2 B^2 \right)
\,\,.
\end{equation}
Another quadratic second--rank tensor is the free electromagnetic
energy--momentum tensor \cite{LL2}
\begin{equation}
\label{tdef}
T^F_{\mu\nu} = 
\varepsilon_0 F_{\mu\alpha} g^{\alpha\beta} F_{\beta\nu} -
{\mathscr L}_F g_{\mu\nu}
\,\,,
\end{equation}
or, in local--galilean coordinates,
\begin{equation}
T^F_{\mu\nu} = 
\left(
\begin{array}{cc}
I & -{\bf S}/c \\
-{\bf S}/c & \sigma
\end{array}
\right)
\,\,,\,\,
T_F^{\mu\nu} = 
\left(
\begin{array}{cc}
I & {\bf S}/c \\
{\bf S}/c & \sigma
\end{array}
\right)
\end{equation}
with 
\begin{eqnarray}
I &=& \frac{\varepsilon_0}{2}\,\left( E^2 + c^2 B^2 \right)
\quad,\quad
{\bf S} =\varepsilon_0 c^2\, {\bf E} \wedge {\bf B}
\,\,,
\nonumber\\
\sigma &=&  
\varepsilon_0 \left[
\left(\frac{E^2}{2} + \frac{c^2 B^2}{2}\right){\bf 1} -
{\bf E} \otimes {\bf E} - c^2 {\bf B} \otimes {\bf B}
\right]
\,\,.
\end{eqnarray}
Here $I$ denotes the intensity, ${\bf S}$ is the Poynting vector,
and $\sigma$ is Maxwell's stress tensor.
The symbols $\wedge$ and $\otimes$ denote the three--dimensional 
vector and tensor product, respectively.

We can only form second--rank tensors from 
$F_{\alpha\beta} \, F^{\alpha'\beta'}$ by some contraction.
Consequently, the linear combinations of the two elementary tensors
${\mathscr L}_F g_{\mu\nu}$ and $T^F_{\mu\nu}$ form the complete class
of second--rank tensors that are quadratic in the field strengths
$F_{\mu\nu}$.

\section{Classical atoms}

\subsection{Postulates}

Consider a classical atom in an electromagnetic field.
The atom is point--like, has a mass $m$, and can sustain induced
electric and magnetic dipoles. 
In the restframe of the atom the dipoles respond to the square of the
electric field strength, $E^2$, and to the magnetic $B^2$,
respectively. 
How does a dielectric atom experience the electromagnetic field when
the atom is moving?

Let us postulate that the atom sees the field as an effective metric. 
Consequently, according to general relativity \cite{LL2}, the action
$S_0$ of the atom is
\begin{equation}
\label{s0}
S_0 = -mc \int ds
\quad,\quad
ds^2 = g^A_{\mu\nu} dx^\mu dx^\nu
\,\,.
\end{equation}
Let us further postulate that the metric of the atom, $g^A_{\mu\nu}$,
is quadratic in the electromagnetic field strengths. 
Any metric is a second--rank tensor.
Hence, we obtain from Sec.\ IIB the general form (\ref{metric})
mentioned in the Introduction.

\subsection{Properties}

A metric of the structure (\ref{metric}) has nice mathematical properties. 
In particular, the contravariant metric tensor $g_A^{\mu\nu}$ (the
inverse of $g^A_{\mu\nu}$) takes on a simple analytic expression,
\begin{equation}
\label{contra_metric}
\sqrt{-g_A} \, g_A^{\mu\nu} =
\sqrt{-g}\left[\left( 1 - a\, {\mathscr L}_F \right) 
g^{\mu\nu} + b\,T_F^{\mu\nu}\right]
\end{equation}
with
\begin{equation}
g_A \equiv {\rm det} (g^A_{\mu\nu})
\end{equation}
and
\begin{equation}
\label{gadet}
\sqrt{-g_A} =
\sqrt{-g}\left[\left( 1 - a\, {\mathscr L}_F \right)^2 - 
\frac{b^2}{4}\,T^F_{\alpha\beta} T_F^{\alpha\beta}\right]
\,\,,
\end{equation}
as one verifies in local--galilean coordinates, with the relation
$T^F_{\alpha\beta} T_F^{\alpha\beta}=
\varepsilon_0^2[(E^2-c^2B^2)^2 + 4c^2({\bf E}\cdot{\bf B})^2]$.

\subsection{Non--relativistic limit}

So far, we have not seen how the metric theory (\ref{metric}) and 
(\ref{s0}) is related to the model of a moving induced dipole. 
Let us consider the non--relativistic limit of velocities low compared
with the speed of light.
This limit corresponds to a motion in an inertial frame close to a 
restframe co--moving with the atom.
We also regard the electromagnetic field energy to be weak compared 
with the atomic rest energy $mc^2$.
We neglect any genuine gravitational field, and obtain in cartesian
coordinates 
\begin{eqnarray}
ds &=& 
\sqrt{
\left(1 - a\, {\mathscr L}_F \right)
\left(c^2 dt^2 - d{\bf x}^2\right) - b\, T^F_{\mu\nu} dx^\mu dx^\nu}
\nonumber\\
&\approx&
\sqrt{c^2 dt^2 - d{\bf x}^2 - 
\left(a\, {\mathscr L}_F + b\, T^F_{00}\right) c^2 dt^2}
\nonumber\\
&\approx& 
\left(1 - \frac{v^2}{2c^2} - 
\frac{a\, {\mathscr L}_F + b\, T^F_{00}}{2}\right) c dt
\end{eqnarray}
with ${\bf v} = d{\bf x}/dt$.
Consequently, we can write the action $S_0$ as
\begin{equation}
S_0 = -mc \int ds \approx \int \left( -mc^2 + L_0\right) dt
\end{equation}
with the non--relativistic Lagrangian
\begin{equation}
L_0 = \frac{m}{2}\, v^2 + 
\frac{\alpha_E}{2}\, E^2 + \frac{\alpha_B}{2}\, c^2 B^2
\end{equation}
and
\begin{eqnarray}
\alpha_E = \frac{a+b}{2}\,\varepsilon_0 mc^2
\quad &,& \quad
\alpha_B = \frac{b-a}{2}\,\varepsilon_0 mc^2
\,\,,
\nonumber\\
a = \frac{\alpha_E-\alpha_B}{\varepsilon_0 mc^2}
\quad &,& \quad
b = \frac{\alpha_E+\alpha_B}{\varepsilon_0 mc^2}
\,\,.
\label{parameters}
\end{eqnarray}
The Lagrangian $L_0$ describes indeed a non--relativistic atom with
electric and magnetic polarizibility $\alpha_E$ and $\alpha_B$,
respectively. 
In this way we have verified that the metric theory (\ref{metric}) and
(\ref{s0}) agrees with the physical picture of traveling dipoles and,
simultaneously, we have been able to express the coefficients $a$ and
$b$ of the metric (\ref{metric}) in terms of atomic quantities.

\section{Matter waves}

\subsection{Postulate}

Gordon has shown \cite{Gordon} that an electromagnetic field
experiences dielectric matter as the effective metric (\ref{gordon}).
Here we postulate that also the opposite is true:
A dielectric matter wave sees the electromagnetic field as a metric,
and in particular as the metric (\ref{metric}) that we have motivated 
for traveling dipoles in Sec.\ III. 
We demonstrate the consistency of this idea with Gordon's theory in
Sec.\ V.
Let us model the matter wave as, fittingly, a complex Klein--Gordon
scalar $\psi$ in an effectively curved space--time.
The action $S_A$ of the atom wave $\psi$ is
\begin{equation}
\label{sa}
S_A = \int {\mathscr L}_A \sqrt{-g}\, d^4x 
\end{equation}
in terms of the Klein--Gordon Lagrangian \cite{LL4}
\begin{eqnarray}
{\mathscr L}_A 
&=& 
\sqrt{\frac{g_A}{g}}
\left[
\frac{1}{2m} g_A^{\mu\nu}
\left(-i\hbar\, \partial_\mu\psi^*\right)
\left(i\hbar\, \partial_\nu\psi\right) -
\frac{mc^2}{2}\,\psi^*\psi \right] 
\nonumber\\
&=&
\sqrt{\frac{g_A}{g}}
\left[
\frac{\hbar^2}{2m}(D_A^\mu\psi^*)(D^A_\mu\psi) -
\frac{mc^2}{2}\,\psi^*\psi \right]
\label{la} 
\end{eqnarray}
where we have employed the covariant derivatives $D^A_\mu$ 
with respect to the effective metric (\ref{metric}).
The action (\ref{sa}) is minimal if the matter wave $\psi$ obeys the
Klein--Gordon equation
\begin{equation}
\label{kleingordon0}
D^A_\mu D_A^\mu\, \psi + \frac{m^2c^2}{\hbar^2}\,\psi = 0
\,\,,
\end{equation}
or, written explicitly \cite{LL2},
\begin{equation}
\label{kleingordon}
\frac{1}{\sqrt{-g_A}}\,\partial_\mu\left(\sqrt{-g_A}\,g_A^{\mu\nu}
\partial_\nu \psi \right) + 
\frac{m^2c^2}{\hbar^2}\,\psi = 0
\,\,.
\end{equation}
Equation (\ref{kleingordon}) together with the functions 
(\ref{contra_metric}) and (\ref{gadet}) and the parameters 
(\ref{parameters}) describes how atomic matter waves respond to 
electromagnetic fields.

\subsection{R{\"on}tgen limit}

Let us prove explicitly that the Klein--Gordon Lagrangian (\ref{la})
contains the known light forces in the limit of relatively low
velocities (compared with $c$) and of weak fields (compared with $mc^2$).
We separate from the atomic wave function $\psi$ the notorious rapid
oscillations due to the rest energy $mc^2$ by defining
\begin{equation}
\varphi \equiv \psi\,\exp\left(i\frac{mc^2}{\hbar}\,t\right)
\,\,.
\end{equation}
We neglect gravity and obtain in cartesian coordinates
\begin{eqnarray}
{\mathscr L}_A
&\approx&
\sqrt{-g_A}\left[
\frac{1}{2}\,g_A^{00}\left(mc^2\varphi^*\varphi + 
i\hbar\,\varphi^*\dot{\varphi} - i\hbar\,\dot{\varphi}^*\varphi\right)
\right.
\nonumber\\
& &
+ \frac{i\hbar c}{2}\,
g_A^{0k}\left(\varphi^* \partial_k\varphi - 
\varphi\, \partial_k\varphi^*\right) 
\nonumber\\
& &
\left.
- \frac{\hbar^2}{2m}\,g_A^{kl}(\partial_k\varphi^*)(\partial_l\varphi)
- \frac{mc^2}{2}\,\varphi^*\varphi
\right]
\nonumber\\
&\approx&
\frac{mc^2}{2}\,\varphi^*\varphi \left(1 - a\,{\mathscr L}_F
+ b\,T_F^{00}\right) 
+ \frac{i\hbar}{2}\left(\varphi^*\dot{\varphi}
- \dot{\varphi}^*\varphi\right)
\nonumber\\
& &
+ \frac{i\hbar c}{2}\, b\,T_F^{0k}\left(\varphi^*\partial_k\varphi -
\varphi\, \partial_k\varphi^*\right)
\nonumber\\
& &
- \frac{\hbar^2}{2m}\,(\nabla\varphi^*)\cdot(\nabla\varphi)
- \frac{mc^2}{2}\,\left(1-2a\,{\mathscr L}_F\right)\varphi^*\varphi
\,\,.
\nonumber\\
&=&
\frac{i\hbar}{2}
\left(\varphi^*\dot{\varphi} - \dot{\varphi}^*\varphi\right)
- \frac{\hbar^2}{2m}\,(\nabla\varphi^*)\cdot(\nabla\varphi)
\nonumber\\
& &
+ \left(\frac{\alpha_E}{2}\, E^2 + \frac{\alpha_B}{2}\, c^2 B^2\right)
\varphi^*\varphi
\nonumber\\
& &
+ \frac{\alpha_E + \alpha_B}{2m}\,
\left( {\bf E} \wedge {\bf B} \right) \cdot
i\hbar \left(\varphi^* \nabla\varphi - 
\varphi\, \nabla\varphi^*\right)
\,\,.
\end{eqnarray}
This result agrees with the R{\"o}ntgen Lagrangian of Ref.\
\cite{LProentgen} in the limit of weak fields and, consequently,
describes indeed the known non--resonant light forces including the
R{\"o}ntgen interaction \cite{Rpapers}.

\subsection{Dielectric flow}

Accelerated by light forces, an atomic matter wave will form a
probability current that appears as a dielectric flow. 
Let us calculate the flow from the phase $S$ of the wave function,
\begin{equation}
\psi = |\psi|\,e^{iS}
\,\,.
\end{equation}
We introduce
\begin{equation}
\label{wflow}
w^\mu \equiv -\frac{\hbar}{mc}\, g_A^{\mu\nu} \partial_\nu S
\,\,,
\end{equation}
and obtain from the Klein--Gordon equation (\ref{kleingordon})
the conservation law of the four--dimensional probability current,
\begin{equation}
\label{pc}
D^A_\mu \left(|\psi|^2 w^\mu\right) =
\frac{1}{\sqrt{-g_A}}\,
\partial_\mu\left(\sqrt{-g_A}\,|\psi|^2 w^\mu\right) = 0
\,\,.
\end{equation}
In the absence of electromagnetic forces, $w^\mu$ describes the
local four--velocity of a free matter wave.
In the presence of a field, we introduce the dielectric flow $u^\mu$
by normalizing $w^\mu$ to unity with respect to the back--ground
metric $g_{\mu\nu}$,
\begin{equation}
\label{udef}
u^\mu \equiv \frac{w^\mu}{w}
\quad,\quad
w \equiv \sqrt{g_{\mu\nu} w^\mu w^\nu}
\,\,.
\end{equation}
We define two densities, $\varrho$ and $\rho$, as
\begin{equation}
\label{rho}
\varrho \equiv |\psi|^2 w \sqrt{\frac{g_A}{g}}
\quad,\quad
\rho \equiv \varrho\, w
\,\,.
\end{equation}
We obtain from the conservation law (\ref{pc})
\begin{equation}
\label{flowing}
\frac{1}{\sqrt{-g}}\,
\partial_\mu\left(\sqrt{-g}\,\varrho\,u^\mu\right) = 
D_\mu \left(\varrho\,u^\mu\right) = 0
\,\,.
\end{equation}
Consequently, $\varrho$ is the scalar probability density of the
atomic de--Broglie wave.
For most practical purposes the two densities $\varrho$ and $\rho$
are identical, because $w$ is unity to a very good approximation.
The difference between $\varrho$ and $\rho$ is subtle:
In Sec.\ VE we show that $mc^2\rho$ is the total enthalpy density of
the dielectric matter wave, with the rest--energy density
$mc^2\varrho$ as the lion's share.

\subsection{Hydrodynamic limit}

As has been mentioned, the objective of this paper is the proof that
the metric interaction (\ref{metric}) between matter waves and light
is compatible with the known theory of dielectrics \cite{Gordon,LL8}.
When a matter wave or, more likely, a macroscopic condensate of
identical matter waves reaches the status of a dielectric it behaves
like a quantum fluid.
In this macroscopic limit the de--Broglie density varies over
significantly larger ranges than the de--Broglie wave length (the same
applies to frequencies), and a hydrodynamic approach has become
extremely successful \cite{BEC}.
Let us approximate
\begin{equation}
i\hbar\,\partial_\nu \psi \approx -\psi\,\hbar\partial_\nu S
\,\,.
\end{equation}
We obtain from the Klein--Gordon Lagrangian (\ref{la}) the
hydrodynamic approximation
\begin{equation}
\label{hydro}
{\mathscr L}_A =
\sqrt{\frac{g_A}{g}}\,|\psi|^2
\left[
\frac{\hbar^2}{2m}\, g_A^{\mu\nu}(\partial_\mu S)(\partial_\nu S) -
\frac{mc^2}{2} 
\right] 
\,\,.
\end{equation}
Let us consider the Euler--Lagrange equations derived from the
hydrodynamic Lagrangian (\ref{hydro}). 
We obtain from the $\partial_\mu S$ dependence of ${\mathscr L}_A$
the dielectric flow (\ref{flowing}) and from a variation with respect 
to $|\psi|^2$ the dielectric Hamilton--Jacobi equation
\begin{equation}
g_A^{\mu\nu}(\partial_\mu S)(\partial_\nu S) = 
\frac{m^2c^2}{\hbar^2}
\,\,,
\end{equation}
or, in terms of the four--vector $w^\mu$ of Eq.\ (\ref{wflow}),
\begin{equation}
\label{wnorm}
g^A_{\mu\nu} w^\mu w^\nu = 1 
\,\,.
\end{equation}
In the hydrodynamic limit the $w^\mu$ vector represents a
four--velocity that is normalized with respect to the effective metric
(\ref{metric}). 
We also see that the hydrodynamic Lagrangian (\ref{hydro}) vanishes at
the actual minimum that corresponds to the physical behavior of a
dielectric matter wave.

\section{Quantum dielectrics}

\subsection{Actio et reactio}

In the previous section we considered a dielectric matter wave in a
given electromagnetic field.
Gordon \cite{Gordon} studied the opposite extreme --- an
electromagnetic field in a given dielectric medium.
Let us address here an intermediate regime of {\it actio et reactio}
where light acts on matter as well as matter acts on light. 
Such a physical regime, characterizing a quantum dielectric, occurs
for example when a Bose--Einstein condensate of an alkali vapor
\cite{BEC} interacts non--resonantly with light \cite{LKP}.
If we were able to arrive at Gordon's metric (\ref{gordon}) from our
starting point (\ref{metric}) we were inclined to take this as
evidence that our approach is right.

To include the dynamics of the electromagnetic field we add the
free--field Lagrangian ${\mathscr L}_F$ to the atomic ${\mathscr L}_A$
in hydrodynamic approximation (\ref{hydro}),
\begin{equation}
{\mathscr L} = {\mathscr L}_F + {\mathscr L}_A
\,\,,
\end{equation}
and regard the electromagnetic field as a dynamic object that is
subject to the principle of least action.
We could also easily include other interactions by additional terms in
${\mathscr L}_A$ such as the atomic collisions within a
Bose--Einstein condensate \cite{BEC} by a Gross--Pitaevskii term.
Let us consider the field variation
\begin{eqnarray}
\delta_F {\mathscr L}
&=& 
\delta_F {\mathscr L}_F
+ \sqrt{\frac{g_A}{g}}\,|\psi|^2\, \frac{\hbar^2}{2m}\, 
(\partial_\mu S)(\partial_\nu S)\,\delta_F g_A^{\mu\nu}
\nonumber\\
& &
+ \sqrt{\frac{g}{g_A}}\,{\mathscr L}_A\,
\delta_F \sqrt{\frac{g_A}{g}}
\,\,.
\end{eqnarray}
As has been mentioned in Sec.\ IVD, the atomic Lagrangian 
${\mathscr L}_A$ vanishes at the minimum of the action, in the
hydrodynamic limit.
We utilize that
\begin{equation}
\delta_F g_A^{\mu\nu} = -g_A^{\mu\alpha} g_A^{\nu\beta} 
\delta_F g^A_{\alpha\beta}
\,\,,
\end{equation}
and obtain, using Eqs.\ (\ref{wflow}-\ref{rho}),
\begin{equation}
\label{deltal}
\delta_F {\mathscr L} = 
\delta_F {\mathscr L}_F - 
\frac{mc^2}{2}\,\rho\,u^\alpha u^\beta \delta_F g^A_{\alpha\beta}
\,\,. 
\end{equation}
The variation of the Lagrangian with respect to the field determines
via the Euler--Lagrange equations the field dynamics.
Can we cast $\delta_F{\mathscr L}$ in the role of a dielectric?

\subsection{Effective Lagrangian}

The principal mathematical artifice of this paper is an effective 
Lagrangian that is designed to agree with ${\mathscr L}$ under 
field variations, and that describes a dielectric medium,
\begin{equation}
\label{effdef}
{\mathscr L}_{\rm EFF} \equiv
{\mathscr L}_F + \frac{mc^2}{2}\,\rho\,
\left(g_{\alpha\beta} - g^A_{\alpha\beta}\right) u^\alpha u^\beta
\end{equation}
with
\begin{equation}
\label{eff}
\delta_F {\mathscr L} = 
\delta_F {\mathscr L}_{\rm EFF}
\,\,.
\end{equation}
Note that the two field variations in the relation (\ref{eff}) differ
in a subtle way:
On the left--hand side, $\delta_F$ abbreviates the total variation
with respect to the electromagnetic field, whereas on the right--hand
side of Eq.\ (\ref{eff}) we treat $\varepsilon$, $\mu$, and $u^\alpha$
as being fixed, despite their hidden dependence on the field due to the
relations (\ref{wflow}-\ref{rho}).

We show explicitly in Sec.\ VD that ${\mathscr L}_{\rm EFF}$ is 
indeed the desired Lagrangian of light in a dielectric medium.
Here we note that ${\mathscr L}_{\rm EFF}$ may metamorphose into a
multitude of forms.
For example, we introduce the permittivity $\varepsilon$ and the
magnetic permeability $\mu$ in terms of elementary atomic quantities 
and in accordance with the parameters (\ref{ab}) mentioned in the 
Introduction
\begin{equation}
\label{epsmu}
\varepsilon = 1 + \frac{\alpha_E}{\varepsilon_0}\,\rho
\quad,\quad
\frac{1}{\mu} = 1 - \frac{\alpha_B}{\varepsilon_0}\,\rho
\,\,.
\end{equation}
In this way we obtain directly from Eqs.\ (\ref{metric}) and (\ref{ab})
\begin{equation}
{\mathscr L}_{\rm EFF} =
\frac{1}{2}\left[
\left(\varepsilon + \frac{1}{\mu}\right){\mathscr L}_F +
\left(\varepsilon - \frac{1}{\mu}\right)u^\alpha u^\beta
T^F_{\alpha\beta}
\right]
\,\,.
\end{equation}
We can also express the effective Lagrangian as 
\begin{equation}
{\mathscr L}_{\rm EFF} =
\frac{1}{\mu}\,{\mathscr L}_F +
\varepsilon_0 \frac{\varepsilon\mu-1}{2\mu}\,
F_{\alpha'\beta'} F_{\alpha\beta}\, u^\alpha u^{\alpha'} g^{\beta\beta'}
\,\,,
\end{equation}
due to the definition (\ref{tdef}) of the free--field energy--momentum
tensor, or we may perform further manipulations, utilizing the
relations
\begin{eqnarray}
&F_{\alpha'\beta'}& F_{\alpha\beta}\, 
u^\alpha u^{\alpha'} g^{\beta\beta'}
= F_{\alpha'\beta'} F_{\alpha\beta}\, 
g^{\alpha\alpha'}u^\beta u^{\beta'}
\,\,,
\nonumber\\
&F_{\alpha'\beta'}& F_{\alpha\beta}\, 
u^\alpha u^{\alpha'} u^\beta u^{\beta'} = 0
\,\,,
\end{eqnarray}
due to the symmetry of the back--ground metric $g^{\alpha\beta}$ 
and the anti--symmetry of the field--strength tensor
$F_{\alpha\beta}$.

\subsection{Gordon's metric}

Quite remarkably, one can express the effective Lagrangian in the form
\cite{Gordon}
\begin{equation}
\label{lgordon}
{\mathscr L}_{\rm EFF} =
- \frac{\varepsilon_0}{4\mu}\,F_{\alpha\beta} F^{(\alpha)(\beta)}
\end{equation}
with
\begin{equation}
F^{(\alpha)(\beta)} \equiv
g_F^{\alpha\alpha'} g_F^{\beta\beta'} F_{\alpha'\beta'}
\end{equation}
and
\begin{equation}
\label{gordonup}
g_F^{\alpha\beta} = g^{\alpha\beta} + 
(\varepsilon\mu - 1)\, u^\alpha u^\beta
\,\,.
\end{equation}
The effective Lagrangian appears as the free electromagnetic 
Lagrangian in a curved space--time with metric (\ref{gordonup}). 
A short exercise proves that $g_F^{\alpha\beta}$ is the inverse of 
$g^F_{\alpha\beta}$, i.e., as the notation suggests it, the
contravariant metric tensor with respect to the covariant 
$g^F_{\alpha\beta}$. 
Consequently, we have indeed arrived at Gordon's space--time geometry
of light in moving media, starting from our metric (\ref{metric}),
which supports the validity of our postulates.

Note that Gordon's space--time geometry is not completely perfect
\cite{Gordon}.
The metrics (\ref{gordon}) and (\ref{gordonup}) depend only on the
square of the refractive index, $\varepsilon\mu$, whereas a
dielectric medium is characterized by two dielectric constants
$\varepsilon$ and $\mu$, in general.
What is the imperfection in the Lagrangian (\ref{lgordon})?
In order to describe a density in general relativity, and in particular
a Lagrangian density, we must consider the determinant of the
metric that describes the scaling of space and time.
Gordon \cite{Gordon} calculated the determinant by employing
co--moving medium coordinates, with the result
\begin{equation}
g_F \equiv {\rm det}\left(g^F_{\alpha\beta}\right) 
= \frac{g}{\varepsilon\mu}
\,\,.
\end{equation}
Hence we obtain the effective action
\begin{eqnarray}
S_{\rm EFF} 
&=&
\int {\mathscr L}_{\rm EFF}\, \sqrt{-g}\, d^4x
\nonumber\\ 
&=&
-\frac{\varepsilon_0}{4}
\int \sqrt{\frac{\varepsilon}{\mu}}\,F_{\alpha\beta}
F^{(\alpha)(\beta)}\,
\sqrt{-g_F}\,d^4x
\end{eqnarray}
that may deviate from the perfect
\begin{equation}
S_F =
-\frac{\varepsilon_0}{4}
\int F_{\alpha\beta} F^{(\alpha)(\beta)}\,
\sqrt{-g_F}\,d^4x
\end{equation}
when $\varepsilon/\mu$ varies significantly.
However, when the density profile of the quantum liquid varies smoothly
compared with the wave length of light we can neglect the variation of
$\varepsilon/\mu$.
Ultracold atoms or Bose--Einstein condensates \cite{BEC} are usually
in this regime that is also compatible with the hydrodynamic behavior
of the quantum liquid.

\subsection{Maxwell's equations}

The first group of Maxwell's equations follows from the structure
(\ref{fdef}) of the field--strength tensor $F_{\mu\nu}$.
The Euler--Lagrange equations of the effective Lagrangian 
(\ref{lgordon}) yield the second group \cite{Gordon,LL8},
\begin{equation}
D_\alpha H^{\alpha\beta} = 0
\quad\mbox{or}\quad
\partial_\alpha\left(\sqrt{-g}\,H^{\alpha\beta}\right) = 0
\end{equation}
with the constitutive equations
\begin{equation}
\label{constitution}
H^{\alpha\beta} = \frac{\epsilon_0}{\mu}\, F^{(\alpha)(\beta)}
\,\,.
\end{equation}
In local--galilean coordinates we can represent $H^{\alpha\beta}$ in
terms (\ref{hmn}) of the dielectric ${\bf D}$ and ${\bf H}$ fields in
SI units.
In this way we find yet another physically meaningful expression for
the effective Lagrangian,
\begin{equation}
\label{dielectric}
{\mathscr L}_{\rm EFF} =
- \frac{1}{4}\,F_{\alpha\beta} H^{\alpha\beta}
= \frac{{\bf E}\cdot{\bf D}}{2} -  \frac{{\bf B}\cdot{\bf H}}{2}
\,\,,
\end{equation}
which is indeed the explicit form of the Lagrangian for the
electromagnetic field in a linear dielectric.

Equation (\ref{constitution}) is equivalent \cite{Gordon} to 
Minkowski's constitutive equations in a moving medium 
\cite{LL8,Minkowski}.
In the limit of low velocities we recover the familiar relations
${\bf D} = \varepsilon_0 \varepsilon\,{\bf E}$ and
$\mu{\bf H} = \varepsilon_0 c^2 {\bf B}$, and, via Eq.\ (\ref{epsmu}),
\begin{equation}
\label{dh}
{\bf D} \approx (\varepsilon_0 + \alpha_E\varrho)\, {\bf E}
\quad,\quad
{\bf H} \approx (\varepsilon_0 - \alpha_B\varrho)\, c^2{\bf B}
\,\,,
\end{equation}
assuming a weak field when $\rho\approx\varrho$.
Relativistic first--order corrections lead to the constitutive
equations derived in Ref.\ \cite{LProentgen} that describe, for
example, the R\"ontgen effect \cite{Roentgen} or lead to Fresnel's
light drag \cite{Fresnel} measured in Fizeau's experiment
\cite{Fizeau}.

In case of a smooth dielectric density we can regard $\varepsilon/\mu$
as a constant, and obtain from Maxwell's equations
\begin{equation}
\partial_\alpha\left(\sqrt{-g_F}\,F^{(\alpha)(\beta)}\right) = 0
\quad\mbox{or}\quad
D^F_\alpha F^{(\alpha)(\beta)} = 0
\,\,.
\end{equation}
Light experiences the quantum dielectric as the space--time metric
(\ref{gordon}), i.e. as an effective gravitational field.

\subsection{Energy--momentum tensor}

According to Antoci and Mihich \cite{Antoci} Gordon \cite{Gordon} 
has already settled the notorious debate about Minkowski's 
\cite{Minkowski} versus Abraham's \cite{Abraham} energy--momentum 
tensor in Abraham's favor.
However, in his paper \cite{Gordon}, Gordon assumed the dielectric
properties of the medium $\varepsilon$, $\mu$, and $u^\alpha$, as
preassigned quantities. 
Having done so, the derived energy--momentum tensor is valid if and
only if the dielectric quantities are constants, i.e. in the case of a
uniform medium, because the conservation of energy and momentum
presupposes the homogeneity of space--time, according to Noether's
theorem.
If one tries to determine the energy and momentum of the
electromagnetic field in an inhomogeneous medium one must not consider
the dielectric properties as given functions, but rather as being 
generated by a physical object, such as the quantum dielectric studied
in this paper.
In short, one should take into account {\it actio et reactio},
and in particular the back action of the medium (an effect seen 
experimentally \cite{Ashkin}).
Does Abraham's tensor have significance beyond uniform media?

Let us determine the energy--momentum tensor via the royal road of
general relativity, as a variation of the Lagrangian with respect 
to the back--ground metric \cite{LL2}, 
\begin{equation}
\label{emt}
T^{\mu\nu} = -\frac{2}{\sqrt{-g}}\,
\frac{\delta\left(\sqrt{-g}\,{\mathscr L}\right)}{\delta g_{\mu\nu}}
= -2\,\frac{\delta{\mathscr L}}{\delta g_{\mu\nu}} 
- {\mathscr L}\,g^{\mu\nu}
\,\,.
\end{equation}
A metric variation $\delta_g$ of the Lagrangian gives, in analogy 
with Eq.\ (\ref{deltal}) and the considerations in Sec.\ VB,
\begin{eqnarray}
\delta_g {\mathscr L}
&=&
\delta_g {\mathscr L}_F 
- \frac{mc^2}{2}\,\rho\,u^\alpha u^\beta\,
\delta_g g^A_{\alpha\beta}
\nonumber\\
&=&
\delta_g {\mathscr L}_{\rm EFF} 
- \frac{mc^2}{2}\,\rho\,u^\alpha u^\beta\,
\delta_g g_{\alpha\beta}
\,\,.
\end{eqnarray}
We recall that ${\mathscr L}_A$ vanishes in the hydrodynamic limit.
Consequently, we arrive at the total energy--momentum tensor in the
form
\begin{equation}
T^{\mu\nu} = 
-2\,\frac{\delta{\mathscr L_{\rm EFF}}}{\delta g_{\mu\nu}} 
- {\mathscr L}_F\,g^{\mu\nu}
+ mc^2\rho\,u^\mu u^\nu
\,\,.
\end{equation}
We represent this expression as the sum 
\begin{equation}
T^{\mu\nu} = T_A^{\mu\nu} + T_{\rm EFF}^{\mu\nu}
\end{equation}
with the atomic component
\begin{eqnarray}
\label{atensor}
T_A^{\mu\nu} 
&=& 
mc^2\rho\,u^\mu u^\nu - p\,g^{\mu\nu}
\,\,,\\
p &=& {\mathscr L}_F - {\mathscr L_{\rm EFF}}
= \frac{1}{4}\,F_{\alpha\beta}
\left(H^{\alpha\beta} - \varepsilon_0 F^{\alpha\beta}\right)
\,\,,
\label{pressure}
\end{eqnarray}
and 
\begin{equation}
\label{efftensor}
T_{\rm EFF}^{\mu\nu} =
-2\,\frac{\delta{\mathscr L_{\rm EFF}}}{\delta g_{\mu\nu}} 
- {\mathscr L_{\rm EFF}}\,g^{\mu\nu}
\,\,.
\end{equation}
We are inclined to interpret the tensor (\ref{efftensor}) as the 
effective energy--momentum tensor of the electromagnetic field in the 
presence of a dielectric medium.

The atomic tensor (\ref{atensor}) appears as the energy--momentum of a
fluid under the dielectric pressure (\ref{pressure}).
In the limit of low flow velocities the pressure approaches 
$-\varepsilon_0 (\alpha_E E^2 + \alpha_B c^2 B^2)\varrho/2$,
according to Eqs.\ (\ref{dielectric}) and (\ref{dh}).
In this limit, atomic dipoles with positive $\alpha_E$ and $\alpha_B$
are attracted towards increasing field intensities.
We also see from the atomic energy--momentum tensor (\ref{atensor})
that a dielectric fluid possesses the total enthalpy density
$mc^2\rho = mc^2w\varrho$, including the relativistic rest energy.
In this way we have found an interpretation for the density $\rho$
that appears at the prominent place (\ref{ab}).
To calculate the enthalpy, we express the effective Lagrangian 
(\ref{effdef}) in terms of the norm $w$.
We use the definition (\ref{udef}) of the four--velocity $u^\alpha$
and the normalization (\ref{wnorm}) of the $w^\alpha$, and obtain
\begin{equation}
p = {\mathscr L}_F - {\mathscr L_{\rm EFF}}
= \frac{mc^2\varrho}{2}\,\left(\frac{1}{w} - w\right)
\,\,,
\end{equation}
or, by inversion,
\begin{equation}
\label{enthalpy}
mc^2\rho = mc^2w\varrho = \sqrt{m^2 c^4 \varrho^2 + p^2} - p
\,\,.
\end{equation}
This equation describes how the enthalpy density depends on the
pressure and on the dielectric density.
On the other hand, Eq.\ (\ref{pressure}) quantifies the pressure that 
depends on the dielectric density and flow, and on the electromagnetic
field as an external quantity.
We may interpret the two formulas (\ref{pressure}) and
(\ref{enthalpy}) as the equations of state for the quantum dielectric.
The density of the fluid's internal energy is the difference between
enthalpy density and pressure \cite{LL6}
\begin{equation}
\epsilon = \sqrt{m^2 c^4 \varrho^2+ p^2} - 2p
\,\,.
\end{equation}
We see that the internal energy approaches 
$mc^2 + \varepsilon_0 (\alpha_E E^2 + \alpha_B c^2 B^2)$
in the limit of a slow flow and a low dielectric pressure.
Atomic dipoles with positive $\alpha_E$ and $\alpha_B$ seem to gain 
internal energy in the presence of an electromagnetic field.

Let us turn to the energy--momentum tensor of the field.
The effective Lagrangian ${\mathscr L}_{\rm EFF}$ characterizes a
medium with preassigned dielectric functions $\varepsilon$ and $\mu$,
i.e. Gordon's case \cite{Gordon}.
Consequently \cite{Gordon}, the effective energy--momentum tensor of
the electromagnetic field is Abraham's \cite{Abraham}
\begin{equation}
T_{\rm EFF}^{\mu\nu} = 
T_{\rm Ab}^{\mu\nu} =
T_{\rm Mk}^{\mu\nu} - (\varepsilon\mu - 1)\,u^\mu \Omega^\nu
\,\,,
\end{equation}
with Minkowski's tensor \cite{Minkowski},
\begin{equation}
T_{\rm Mk}^{\mu\nu} =  
H^{\mu\alpha}  F_{\alpha\beta}\, g^{\beta\nu} +
\frac{1}{4}\, H^{\alpha\beta}  F_{\alpha\beta}\,g^{\mu\nu}
\,\,,
\end{equation}
corrected by the {\it Ruhstrahl} \cite{Abraham}
\begin{equation}
\Omega^\nu =
F_{\alpha\alpha'} u^{\alpha'} u_\beta
\left(
H^{\alpha\beta} u^\nu +
H^{\beta\nu} u^\alpha +
H^{\nu\alpha} u^\beta
\right)
\,\,.
\end{equation}
In locally co--moving galilean coordinates or in a medium at rest,
the spatial component of the {\it Ruhstrahl} is 
proportional to the Poynting vector (hence the name), 
\begin{equation}
\Omega^\nu =
\left(0, \frac{{\bf E}\wedge{\bf H}}{c}\right)
\,\,.
\end{equation}
In this case the effective energy--momentum tensor of the field 
takes the form
\begin{equation}
T_{\rm Ab}^{\mu\nu} = 
\left(
\begin{array}{cc}
I & {\bf S}/c \\
{\bf S}/c & \sigma
\end{array}
\right)
\end{equation}
with intensity $I$, Poynting vector ${\bf S}$, and stress tensor
$\sigma$
\begin{eqnarray}
I &=& \frac{{\bf E}\cdot{\bf D}}{2} + \frac{{\bf B}\cdot{\bf H}}{2}
\quad,\quad
{\bf S} ={\bf E} \wedge {\bf H}
\,\,,
\nonumber\\
\sigma &=& 
\left(\frac{{\bf E}\cdot{\bf D}}{2} + \frac{{\bf B}\cdot{\bf H}}{2}
\right){\bf 1} - {\bf E} \otimes {\bf D} - {\bf B} \otimes {\bf H}
\,\,.
\end{eqnarray}
We see that Abraham's tensor describes indeed the effective
energy--momentum of the electromagnetic field, even in the general
case of a non--uniform medium that is able to move under the
pressure of light forces. 

\section{Credo}

Light experiences dielectric matter as an effective gravitational field
\cite{Gordon,PhamMauQuan,LPstor,LPliten} and matter experiences light
as a form of gravity as well. 
Light and matter see each other as dual space--time metrics, a
unique model in field theory, to the knowledge of the author. 
We have solidified this mental picture by postulating the idea and
demonstrating its striking consistency with the theory of dielectrics
\cite{Gordon,LL8}. 
It would be interesting to see whether our model can be derived
directly from first principles. 
In passing, we have determined the energy--momentum tensor that 
governs {\it actio et reactio} of electromagnetic fields in 
quantum dielectrics.
The tensor is Abraham's \cite{Abraham} plus the energy--momentum of
the medium characterized by a dielectric pressure and an enthalpy
density.

Our idea may serve as a guiding line for understanding the effects of
slow light \cite{slowlight} on matter waves.
Here one can conceive of creating light fields that appear to atoms as
quasi--astronomical objects.
The holy grail in this field would be the creation of a black hole
made of light.

Light and matter interact with each other as if both were
gravitational fields, and light and matter are genuine quantum fields 
in Nature. 
A distinct quantum regime of dielectrics has been prepared in the 
laboratories where Bose--Einstein condensates of alkali vapors
\cite{BEC} interact non--resonantly with light quanta, but has never
been viewed as an analogue of quantum gravity, to the knowledge of the
author.
Sound in superfluids \cite{Supersound} and in alkali Bose--Einstein 
condensates \cite{Garay} has been considered as a quantum field in a 
curved space--time, as being able to emit the acoustic analogue of 
Hawking radiation \cite{Hawking}. 
However, the quantum sound still propagates in a classical medium, 
in contrast to light quanta in a quantum dielectric.
In many respects, we have reasons to hope that Bose--Einstein 
condensates may serve as testable prototype models for quantum gravity.

\section*{Acknowledgements}

I am very grateful to 
Sir Michael Berry,
Ignacio Cirac,
Carsten Henkel,
Susanne Klein,
Rodney Loudon,
Paul Piwnicki,
Stig Stenholm,
and
Martin Wilkens
for conversations on moving media.
I acknowledge the generous support of 
the Alexander von Humboldt Foundation
and of the G\"oran Gustafsson Stiftelse.



\begin{thebibliography}{99}

\bibitem{Gordon}
W. Gordon, Ann. Phys. (Leipzig) {\bf 72}, 421 (1923).

\bibitem{PhamMauQuan}
Pham Mau Quan,
C. R. Acad. Sci. (Paris) {\bf 242}, 465 (1956);
Archive for Rational Mechanics and Analysis {\bf 1}, 54 (1957/58).

\bibitem{LPstor}
U. Leonhardt and P. Piwnicki,
Phys. Rev. A {\bf 60}, 4301 (1999).

\bibitem{LPliten}
U. Leonhardt and P. Piwnicki,
Phys. Rev. Lett. {\bf 84}, 822 (2000).

\bibitem{Moeller}
C. M{\o}ller,
{\it The Theory of Relativity}
(Oxford University Press, Oxford, 1972).

\bibitem{Fresnel}
A. J. Fresnel,
Ann. Chim. Phys. {\bf 9}, 57 (1818).

\bibitem{Fizeau}
H. Fizeau,
C. R. Acad. Sci. (Paris) {\bf 33}, 349 (1851).

\bibitem{BornWolf}
M. Born and E. Wolf,
{\it Principles of Optics}
(Cambridge University Press, Cambridge, 1999).

\bibitem{Bortolotti}
E. Bortolotti, 
Rend. R. Acc. Naz. Linc., 6a,  {\bf 4}, 552 (1926).

\bibitem{atom_optics}
See e.g. 
P. Berman (ed.),
{\it Atom Interferometry}
(Academic, San Diego, 1997).

\bibitem{LL2}
L. D. Landau and E. M. Lifshitz,
{\it The Classical Theory of Fields}
(Pergamon, Oxford, 1975).

\bibitem{Rpapers}
H. Wei, R. Han, and X. Wei,
Phys. Rev. Lett. {\bf 75}, 2071 (1995),
see also Refs.\ \cite{LProentgen} and \cite{Roentgen}, and
M. Babiker, E. A. Power, and T. Thirunamachandran,
Proc. Roy. Soc. A {\bf 332}, 187 (1973);
M. Babiker,
J. Phys. B {\bf 17}, 4877 (1984);
C. Baxter, M. Babiker, and R. Loudon,
Phys. Rev. A {\bf 47}, 1278 (1993);
V. Lembessis, M. Babiker, C. Baxter, and R. Loudon,
{\it ibid.} {\bf 48}, 1594 (1993);
M. Wilkens,
Phys. Rev. A {\bf 49}, 570 (1994);  
Phys. Rev. Lett. {\bf 72}, 5 (1994);
{\it ibid.} {\bf 81}, 1533 (1998);
G. Spavieri,
{\it ibid.} {\bf 81}, 1533 (1998);
{\bf 82}, 3932 (1999);
Phys. Rev. A {\bf 59}, 3194 (1999).

\bibitem{slowlight}
L. V. Hau, S. E. Harris, Z. Dutton, and C. H. Behroozi,
Nature {\bf 397}, 594 (1999);
M. M. Kash, V. A. Sautenkov, A. S. Zibrov, L. Hollberg,
G. R. Welch, M. D. Lukin, Y. Rostovsev, E. S. Fry, and M. O. Scully,
Phys. Rev. Lett. {\bf 82}, 5229 (1999);
D. Budiker, D. F. Kimball, S. M. Rochester, and V. V. Yashchuk,
{\it ibid.} {\bf 83}, 1767 (1999).

\bibitem{LL8}
L. D. Landau and E. M. Lifshitz,
{\it Electrodynamics of Continuous Media}
(Pergamon, Oxford, 1984).

\bibitem{LL4}
L. D. Landau and E. M. Lifshitz,
{\it Quantum Electrodynamics}
(Pergamon, Oxford, 1982).

\bibitem{LProentgen}
U. Leonhardt and P. Piwnicki,
Phys. Rev. Lett. {\bf 82}, 2426 (1999).

\bibitem{BEC}
F. Dalfovo, S. Giorgini, L. P. Pitaevskii, and S. Stringari,
Rev. Mod. Phys. {\bf 71}, 463 (1999).

\bibitem{LKP}
U. Leonhardt, T. Kiss, and P. Piwnicki, 
Euro. Phys. J. D {\bf 7}, 413 (1999). 

\bibitem{Minkowski}
H. Minkowski,
Nachr. d. K. Ges. d. Wiss. zu G\"ott., Math.-Phys. Kl. 53 (1908).

\bibitem{Roentgen}
W. C. R\"ontgen,
Ann. Phys. Chem. {\bf 35}, 264 (1888).

\bibitem{Antoci}
S. Antoci and L. Mihich,
Nuovo Cimento B {\bf 112}, 991 (1997);
Euro. Phys. J. D {\bf 3}, 205 (1998).

\bibitem{Abraham}
M. Abraham,
Rend. Circ. Matem. Palermo {\bf 28}, 1 (1909); {\bf 30}, 33 (1910).

\bibitem{Ashkin}
A. Ashkin and J. M. Dziedzic, 
Phys. Rev. Lett. {\bf 30}, 139 (1973). 

\bibitem{LL6}
L. D. Landau and E. M. Lifshitz,
{\it Fluid Mechanics}
(Pergamon, Oxford, 1987).

\bibitem{Supersound}
W. G. Unruh,
Phys. Rev. Lett. {\bf 46}, 1351 (1981);
Phys. Rev. D {\bf 51}, 2827 (1995);
T. A. Jacobson,
{\it ibid.} {\bf 44}, 1731 (1991);
T. A. Jacobson and G. E. Volovik,
{\it ibid.} {\bf 58}, 064021 (1998);
N. B. Kopnin and G. E. Volovik,
JETP Lett. {\bf 67}, 140 (1998);
T. A. Jacobson and G. E. Volovik,
{\it ibid.}, {\bf 68}, 874 (1998);
G. E. Volovik,
{\it ibid.}, {\bf 69}, 705, (1999);
M. Visser, 
Class. Quantum Grav. {\bf 15}, 1767 (1998).

\bibitem{Garay}
L. J. Garay, J. R. Anglin, J. I. Cirac, and P. Zoller,
arXiv:gr-qc/0002015.

\bibitem{Hawking}
S. M. Hawking, 
Nature {\bf 248}, 30 (1974); 
Commun. math. Phys. {\bf 43}, 199 (1975).

\end{thebibliography}
\end{document}